\let\orilabel\label
\documentclass[aps,prmaterials,reprint,superscriptaddress,floatfix]{revtex4-2}
\let\label\orilabel
\usepackage{graphicx}
\usepackage{amsmath,amssymb}
\usepackage{hyperref}
\usepackage{enumitem}
\usepackage{booktabs}
\AtBeginDocument{%
  \setlength{\parindent}{0pt}
  \setlength{\parskip}{0.5\baselineskip}
}
\begin{document}

\title{A depth resolved investigation of hydrogen uptake in carbon based nanostructures by soft-to-hard photoemission spectroscopy}

\author{Orlando Castellano}
\affiliation{Dipartimento di Scienze, Università degli Studi di Roma Tre, Via della Vasca Navale 84, 00146 Roma, Italy}
\affiliation{INFN Sezione di Roma Tre, Via della Vasca Navale 84, 00146 Roma, Italy}

\author{Alice Apponi}
\affiliation{INFN Sezione di Roma Tre, Via della Vasca Navale 84, 00146 Roma, Italy}

\author{Luca Cecchini}
\affiliation{INFN Sezione di Roma, Piazzale Aldo Moro 5, 00185 Roma, Italy}

\author{Daniele Paoloni}
\affiliation{Dipartimento di Scienze, Università degli Studi di Roma Tre, Via della Vasca Navale 84, 00146 Roma, Italy}

\author{Simone Ritarossi}
\affiliation{Dipartimento di Scienze, Università degli Studi di Roma Tre, Via della Vasca Navale 84, 00146 Roma, Italy}
\affiliation{INFN Sezione di Roma Tre, Via della Vasca Navale 84, 00146 Roma, Italy}

\author{Francesco Pandolfi}
\affiliation{INFN Sezione di Roma, Piazzale Aldo Moro 5, 00185 Roma, Italy}

\author{Ilaria Rago}
\affiliation{INFN Sezione di Roma, Piazzale Aldo Moro 5, 00185 Roma, Italy}

\author{Tien-Lin Lee}
\affiliation{Diamond Light Source Ltd, Didcot OX11 0DE, United Kingdom}

\author{Samuel Jeong}
\affiliation{Department of Applied Physics, Institute of Pure and Applied Sciences, University of Tsukuba, 1-1-1 Tennodai, Tsukuba, Ibaraki, 305-8573, Japan}

\author{Yoshikazu Ito}
\affiliation{Department of Applied Physics, Institute of Pure and Applied Sciences, University of Tsukuba, 1-1-1 Tennodai, Tsukuba, Ibaraki, 305-8573, Japan}
\affiliation{Tsukuba Institute for Advanced Research (TIAR), University of Tsukuba, 1-1-1 Tennodai, Tsukuba, Ibaraki, 305-8577, Japan}

\author{Carlo Mariani}
\affiliation{INFN Sezione di Roma, Piazzale Aldo Moro 5, 00185 Roma, Italy}
\affiliation{Sapienza Università di Roma, Piazzale Aldo Moro 5, 00185 Roma, Italy}

\author{Gianluca Cavoto}
\affiliation{INFN Sezione di Roma, Piazzale Aldo Moro 5, 00185 Roma, Italy}
\affiliation{Sapienza Università di Roma, Piazzale Aldo Moro 5, 00185 Roma, Italy}

\author{Francesco Offi}
\affiliation{Dipartimento di Scienze, Università degli Studi di Roma Tre, Via della Vasca Navale 84, 00146 Roma, Italy}
\affiliation{INFN Sezione di Roma Tre, Via della Vasca Navale 84, 00146 Roma, Italy}

\author{Alessandro Ruocco}
\affiliation{Dipartimento di Scienze, Università degli Studi di Roma Tre, Via della Vasca Navale 84, 00146 Roma, Italy}
\affiliation{INFN Sezione di Roma Tre, Via della Vasca Navale 84, 00146 Roma, Italy}

\begin{abstract}
\noindent
Hydrogen chemisorption on graphitic carbon modifies the carbon orbital hybridization from sp² to sp³, altering both structural and electronic properties. Understanding not only the lateral extent but also the depth distribution of hydrogen uptake in three-dimensional carbon architectures is essential for both fundamental studies and storage applications. To this end, we investigate here the evolution of the C 1s core-level lineshape in nanoporous graphene (NPG) and vertically aligned carbon nanotubes (CNTs) upon hydrogenation, exploiting soft-to-hard X-ray photoemission spectroscopy to achieve a depth-resolved analysis. Decomposition of the C 1s spectra reveals the formation of an sp³ rich overlayer, indicating hydrogen chemisorption limited to the outermost accessible surfaces in both systems. These results clarify the depth distribution of hydrogen in curved and porous graphitic networks and provide quantitative constraints on its chemisorption for carbon-based hydrogen storage applications.
\end{abstract}

\maketitle

\section{Introduction}
Hydrogenated graphene is a widely used model system for controllable chemical functionalization of planar carbon networks. It consists in the formation of C–H bonds on the carbon lattice, profoundly modifying its electronic and structural properties. In the fully hydrogenated limit, theory predicts the formation of graphane, a stable insulating phase with a sizeable band gap \cite{sofo_graphane_2007}. Partial and reversible hydrogenation has been experimentally demonstrated by exposure to atomic hydrogen via thermal cracking, hydrogen plasma, or electrochemical reactions \cite{whitener_review_2018}. The resulting material shows a strong suppression of conductivity followed by recovery of the pristine electronic properties upon annealing due to hydrogen desorption \cite{elias_control_2009, balog_controlling_2013}.
\par
Beyond monolayer graphene, three-dimensional graphene-derived nanostructures offer the opportunity to investigate hydrogen chemisorption in the presence of finite thickness, curvature, and internal surfaces. At the same time, they raise a central and still open question: whether hydrogen chemisorption remains confined to the outermost accessible surfaces or can penetrate into the inner layers. Addressing this question is essential for understanding both the fundamental mechanisms of hydrogen adsorption and the practical limits of its uptake in three-dimensional carbon architectures. In this respect carbon nanotubes (CNTs) have been widely investigated for hydrogen chemisorption. Experimental studies, conducted with high surface sensitivity, have demonstrated substantial C–H bond formation upon exposure to atomic hydrogen (or deuterium) in both single-wall CNTs (SWCNTs) \cite{nikitin_hydrogenation_2005,nikitin_hydrogen_2008} and multi-wall CNTs (MWCNTs) \cite{tayyab_atomic_2025,pekker_hydrogenation_2001, gorodetskiy_hydrogen_2020}. Hydrogenation has also been investigated in nanoporous graphene (NPG), a bicontinuous graphene network with nanoscale pores, combining high surface accessibility with a high density of curvature and topological defects. Recent experiments have reported two-sided hydrogenation of graphene in NPG with high surface homogeneity, where modification of the electronic structure leads to band-gap opening \cite{betti_gap_2022, betti_dielectric_2023, betti_homogeneous_2022}.
\par
Understanding hydrogen adsorption in three-dimensional graphene-based materials is also relevant for applications requiring the storage of hydrogen or its isotopes in carbon nanostructures. In experiments such as PTOLEMY \cite{apponi_implementation_2022, betti_neutrino_2019}, graphene-based materials are foreseen as solid-state tritium targets for neutrino-mass measurements and cosmic neutrino background detection. More recently, hydrogenated vertically aligned CNTs arrays have also been proposed as directional sub-GeV dark matter detectors, where the signal would arise from the ejection of chemisorbed hydrogen nuclei \cite{arias_darkmatter_2026}. In both cases, the achievable target mass depends critically on whether hydrogenation is confined to external surfaces or can extend throughout the internal volume of the material. Surface-confined functionalization would severely limit the effective uptake, whereas homogeneous adsorption would grant access to a substantially larger fraction of available carbon sites. Closely related to this issue is the permability of graphene-based nanostructures to atomic hydrogen, a process that remains largely unexplored.
\par
In pristine graphene, each carbon atom is sp² hybridized and participates in a delocalized $\pi$ network that defines the characteristic electronic structure of the material. Upon hydrogen chemisorption, the bonded carbon atom is displaced out of the graphene plane, inducing distortions of the surrounding C–C bonds, corresponding to a local sp³ hybridization \cite{boukhvalov_hydrogen_2008}. The energy cost of C–H bond formation is strongly influenced by the local atomic geometry: curvature, ripples, and lattice defects introduce partial sp³ character and lower the energy barrier for C–H bond formation \cite{ruffieux_hydrogen_2002}. Despite their three-dimensional morphology, CNTs and NPG retain a locally planar sp² bonding framework in which curvature and topological disorder are intrinsic \cite{kleiner_curvature_2001, di_bernardo_topology_2018, di_bernardo_two-dimensional_2017}, potentially enhancing hydrogen chemisorption probability compared to flat graphene \cite{abdelnabi_towards_2021}.
\par
The local transition from sp² to sp³ hybridization and the associated changes in electronic properties induced by hydrogen chemisorption can be probed experimentally through the analysis of the carbon core-level lineshapes. X-ray photoemission spectroscopy (XPS), although intrinsically insensitive to hydrogen due to the absence of core electrons, provides indirect access to C–H bond formation via chemical shifts of the C 1s core level associated with different carbon bonding configurations. In particular, in hydrogenated graphene-based systems, the emergence of an sp³ component is typically reflected in a shift of the C 1s peak toward higher binding energy and in a modification of the line shape, allowing the relative sp² and sp³ contributions to be disentangled \cite{apponi_highly_2025, abdelnabi_towards_2021, betti_gap_2022, tayyab_atomic_2025}. However, conventional XPS, using Al K$\alpha$ (1486.7 eV) excitation or lower energies, remains inherently surface sensitive, with a probing depth limited to a few nanometers by the inelastic mean free path of the emitted photoelectrons \cite{tanuma_calculation_2003}.
\par
Extending the probing depth requires increasing the photon energy into the hard X-ray regime, where the inelastic mean free path is significantly enhanced \cite{sacchi_quantifying_2005}. Hard X-ray photoemission spectroscopy (HAXPES), when combined with soft X-ray measurements, provides depth-sensitive access to the electronic structure and chemical environment of three-dimensional materials. At high photon energies, however, photoemission recoil effects become increasingly relevant, particularly for low-Z elements, leading to systematic energy shifts and additional broadening of core-level features. In a semiclassical framework, the overall spectral modifications associated with recoil can be approximated by a Gaussian contribution \cite{takata_recoil_2007,fujikawa_2006}. In the present work, we adopt this phenomenological approach by allowing for a photon-energy-dependent rigid shift of the C 1s binding energy and by including an additional Gaussian broadening term in the lineshape.
\par
We employ photoemission spectroscopy over a broad photon-energy range, spanning from soft to hard X-rays, to investigate the depth dependence of hydrogen chemisorption in two representative three-dimensional graphene-based nanostructures: multi-walled CNTs and NPG. Hydrogenation is achieved via plasma exposure for CNTs and via thermal cracking for NPG. By systematically comparing hydrogenated samples with their pristine counterparts at different photon energies, we aim to identify and discuss spectral changes consistent with hydrogen-induced sp³ bonding, while considering the concurrent influence of recoil effects intrinsic to high-energy photoemission. To quantify the depth distribution of hydrogenation, we employ a modified thin-overlayer attenuation model for photoemission intensity based on the formalism introduced by Fadley \cite{Fadley1978}.
\section{Sample Preparation and Experimental Methods}
NPG was synthesized by chemical vapor deposition (CVD) on a nanoporous Ni as a CVD template. The resulting three-dimensional graphene network replicated the surface morphology of the Ni-based CVD substrate, which was subsequently removed by chemical dissolution to isolate the graphene, as detailed elsewhere \cite{ito_highquality_2014, ito_multifunctional_2015}. Two samples with the sheet thickness of about 30 um from the same batch were prepared: one pristine and one hydrogenated. Hydrogenation was performed in an ultra high vacuum (UHV) chamber at the LASEC Laboratory (Roma Tre University) using a FOCUS EFM-H atomic hydrogen source, with a partial pressure of $3.6 \times 10^{-6}$ mbar, a heating power of 25 W (capillary temperature $\sim2100$ K), and a total exposure time of 30 h.
\par
Vertically aligned CNTs were grown on Si chips (10 × 10 mm², 500 µm thick) by CVD of acetylene ($\mathrm{C_2H_2}$) at the INFN TITAN facility (Sapienza University of Rome) using a custom-built reactor; optimized growth parameters are reported elsewhere \cite{schifano_plasma-etched_2023, cecchini_quantitative_2025}. The as-grown sample was cut into two halves (approximately 5 × 10 mm²): one half was exposed to a hydrogen plasma (100 W, 0.7 mbar $\mathrm{H_2}$, 300 sccm, 1 h), while the other half was retained as a pristine reference.
\par
Initial XPS measurements to characterize the samples hydrogenation process were performed at the LASEC Laboratory (Roma Tre University). For NPG, XPS measurements were carried out \textit{in situ} during and after completion of hydrogen exposure. C 1s spectra were acquired using a monochromatized Al K$\alpha$ X-ray source ($h\nu = 1486.7$ eV) and a hemispherical electron analyzer. The total energy resolution was 460 meV. The binding energy scale was calibrated using a clean highly oriented pyrolytic graphite (HOPG) reference by setting the C 1s binding energy to 284.5 eV \cite{apponi_highly_2025}.
\par
Soft and hard X-ray photoemission spectroscopy measurements were performed at beamline I09 of Diamond Light Source (UK) \cite{lee_two-color_2018}. During transport, samples were kept under low vacuum and introduced into the analysis chamber after brief air exposure ($\sim15$ min) in order to minimize oxygen–hydrogen substitution effects observed upon prolonged exposure to ambient conditions \cite{apponi_stability_2026}. The samples were subsequently annealed \textit{in situ} under UHV to remove adsorbed contaminants. Hydrogenated NPG (HNPG) and hydrogenated CNTs (HCNTs) samples were annealed at milder temperatures $\leq 400\,^\circ\mathrm{C}$ to avoid hydrogen desorption.
\par
Photoemission spectra of the C 1s core level were collected at photon energies of 0.8, 1.4, 4.1, and 8.1 keV, while O 1s spectra were collected at 1.4, 4.1, and 8.1 keV. Spectra were recorded in normal emission geometry. The binding energy scale was calibrated to the Fermi edge of a polycrystalline Au foil in thermal and electrical contact with the samples. The total energy resolution was approximately 0.30 eV for photon energies between 1.4 and 8.1 keV and 0.15 eV at 0.8 keV.
\par
For each class of samples (NPG and CNT) a global fitting analysis, better described elsewhere \cite{PAOLONI2023122322, apponi_highly_2025}, was performed on the C 1s core-level spectra acquired at the four photon energies. A single global fit was applied to spectra from both pristine and hydrogenated samples. The fitted components, discussed in more detail in the Results and Discussion section, represent distinct chemical environments of carbon, including graphitic sp² carbon, sp³ hybridized carbon, carbon oxide species (C–O and O–C–O), defect-related carbon states, and $\pi$-plasmon loss features.The sp² components were fitted using asymmetric Doniach–Šunjić line shapes convoluted with a Gaussian, whereas all other contributions were modeled using Voigt profiles. A Shirley background was included in all fits.
\par
In the global fitting procedure the binding energy of the sp² component was treated as a free parameter for each spectrum. This can capture both the average recoil shift at high photon energies and possible sample charging in HNPG and HCNTs, due to the hydrogenation-induced reduced conductivity. In contrast, the relative binding energy separations of all other components with respect to the sp² peak were constrained as global parameters shared by all 8 spectra (4 different photon energies for pristine and hydrogenated samples) . The Doniach–Šunjić asymmetry parameter and the Lorentzian widths of each component were also constrained globally. The Gaussian width of each component is a free parameter which takes into account the contributions from the instrumental resolution, recoil broadening, and an additional broadening due to the possible presence of unresolved subcomponents at slightly different binding energies.
\section{Results and Discussion}
C 1s photoemission spectra of both pristine and hydrogenated NPG and CNTs were initially acquired with a laboratory-based setup (monochromatized Al K$\alpha$ X-ray source (h$\nu$ = 1486.7 eV) to verify the effectiveness of the hydrogenation process. Following the approach by Betti et al. \cite{betti_gap_2022}, from the intensities of sp² and sp³ components of the C 1s spectra we can estimate the hydrogen-induced spectral modifications: $sp^3/(sp^3 + sp^2)$ results in a value close to $30\%$ in the case of HNPG and $67\%$ for HCNTs. Hydrogenation of CNTs was performed at a separate facility, followed by \textit{ex situ} characterization. In contrast, NPG was hydrogenated \textit{in situ}, enabling XPS measurements on the same sample before and after exposure under identical conditions. For synchrotron measurements, pristine and hydrogenated samples from the same growth batch were compared.
\par
\begin{figure}[htbp]
    \centering            
    \includegraphics[width=\linewidth]{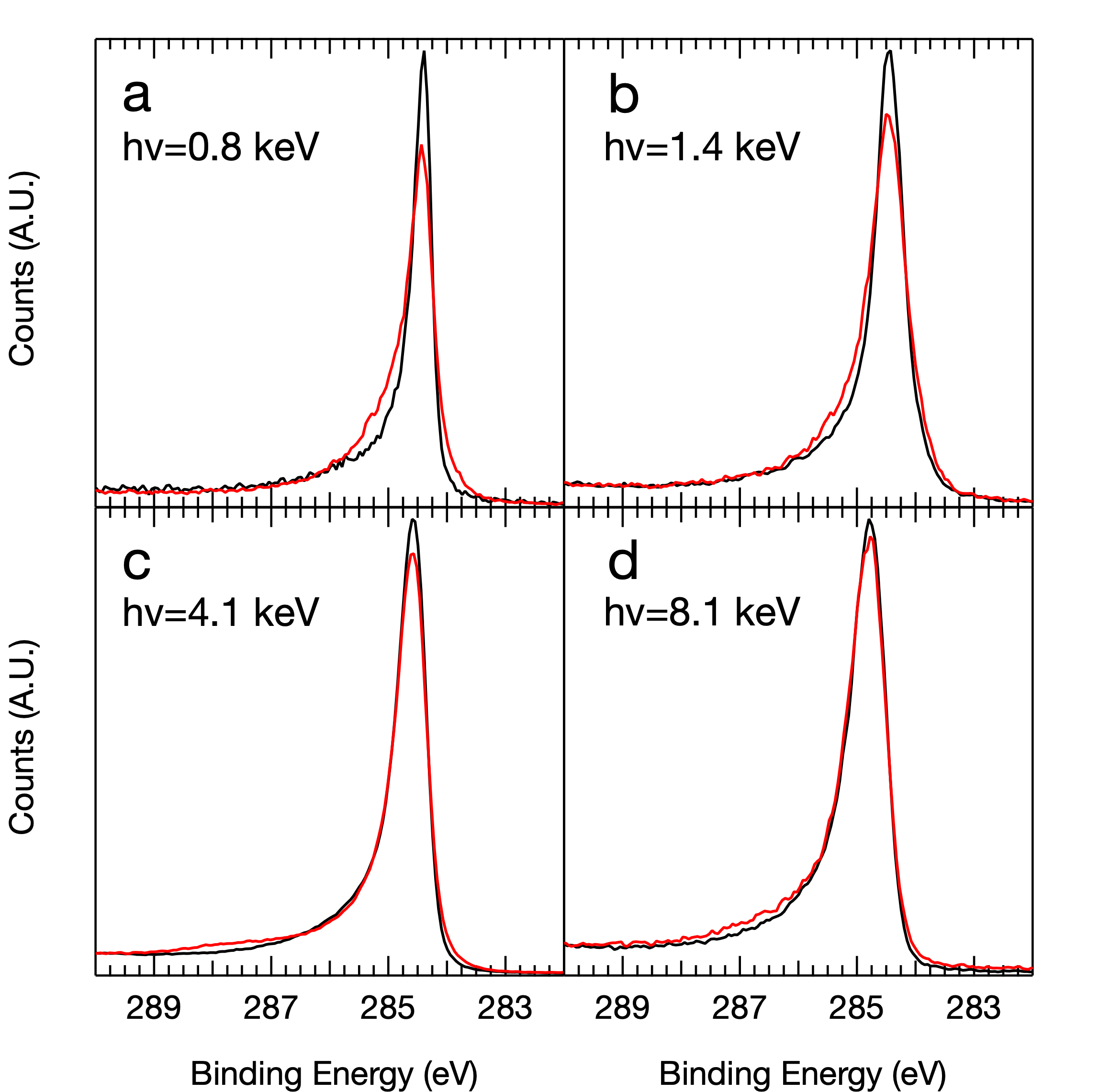}
        \caption{C 1s spectra for NPG (black) and HNPG (red). Spectra are taken in normal emission at different photon energies: 0.8 keV (a), 1.4 keV (b), 4.1 keV (c), 8.1 keV (d) and normalized to the total area.}
        \label{fig:1}
    \hfill
        \centering
    \includegraphics[width=\linewidth]{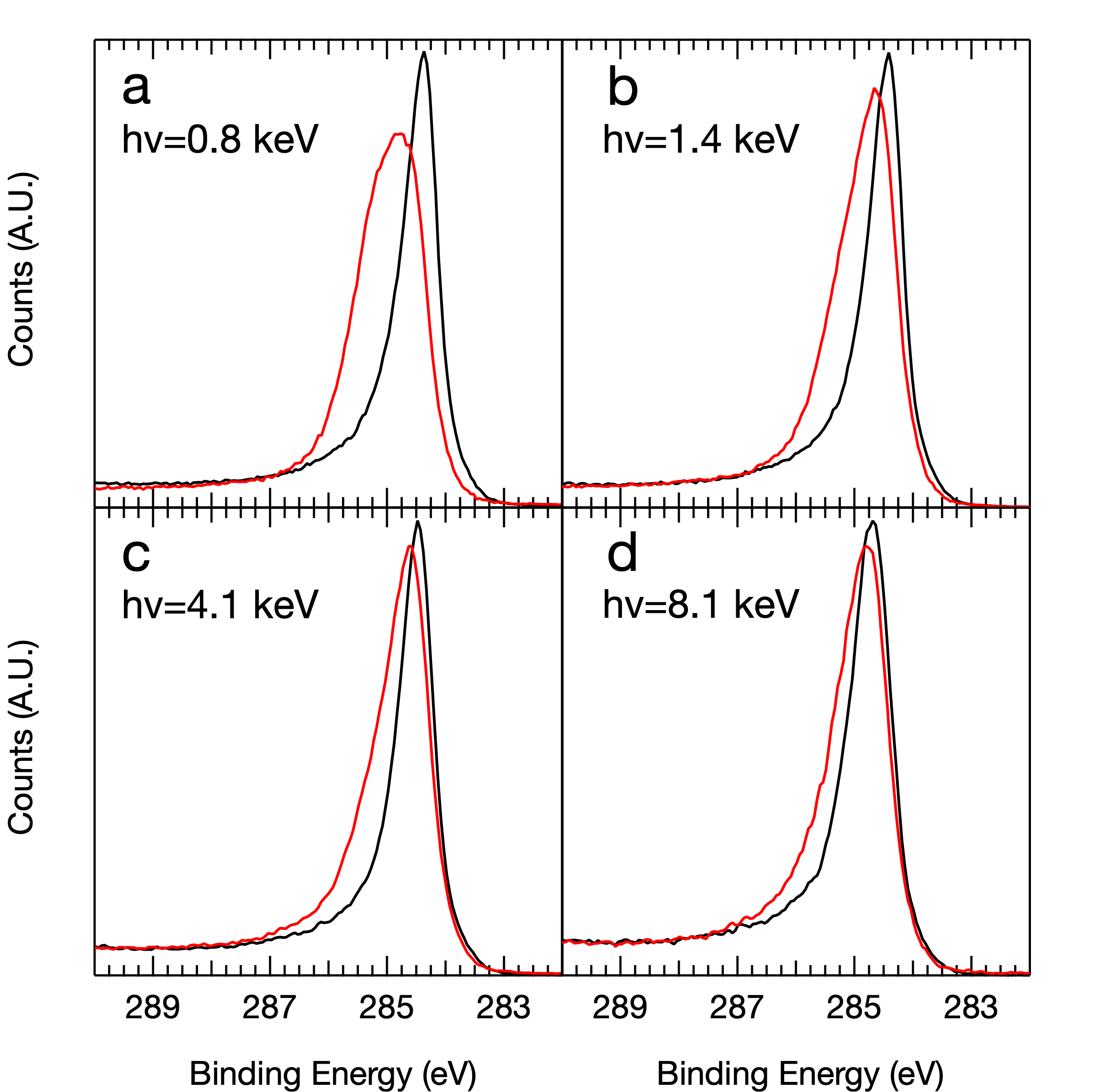}
        \caption{C 1s spectra for CNTs (black) and HCNTs (red). Spectra are taken in normal emission at different photon energies: 0.8 keV (a), 1.4 keV (b), 4.1 keV (c), 8.1 keV (d) and normalized to the total area.}
        \label{fig:2}
\end{figure}

Figure \ref{fig:1} shows the C 1s spectra of pristine NPG (black) and hydrogenated HNPG (red) acquired at photon energies of 0.8 keV (a), 1.4 keV (b), 4.1 keV (c), and 8.1 keV (d) in normal emission. Intensities are normalized to the total area and binding energies are referenced to the measured Fermi level. At 0.8 keV photon energy, corresponding to the highest surface sensitivity, the spectra differ markedly: the HNPG spectrum exhibits a broader main peak with a pronounced tail toward higher binding energies with respect to NPG. With increasing photon energy and probing depth, the spectral differences progressively diminish: at 1.4 keV they are already less pronounced, while at 4.1 keV and 8.1 keV the HNPG and NPG spectra appear nearly indistinguishable. The absence of significant differences at high photon energy already suggests that hydrogen induced modifications of the lineshape do not extend uniformly throughout the probed sample thickness.
\par
Similarly, Figure \ref{fig:2} shows the C 1s spectra of pristine CNTs (black) and HCNTs (red) acquired at the same photon energies. For all excitation energies, clear differences between HCNTs and CNTs are observed. Compared to CNTs, the HCNTs spectra display increased linewidth, a slight shift of the main peak toward higher binding energy, and an enhanced high binding energy tail. These lineshape differences gradually decrease at higher photon energies, indicating a reduced hydrogen presence at larger probing depths. However, even at the highest photon energy of 8.1 keV, a measurable difference between pristine and hydrogenated CNTs remains, suggesting that hydrogen-induced modifications are not limited to the outermost probed region, although they remain non-uniform in depth.
\par
A direct quantitative comparison between HNPG and HCNTs should be treated cautiously, since the two systems were hydrogenated by different methods. The larger sp³ fraction observed in HCNTs may therefore reflect not only differences in morphology and accessible surface area, but also differences in hydrogenation efficiency associated with the preparation route. Nevertheless, previous studies have reported hydrogen or deuterium coverages in carbon nanotubes comparable to those observed here also using thermal-cracking sources rather than plasma-based hydrogenation \cite{nikitin_hydrogenation_2005, nikitin_hydrogen_2008, tayyab_atomic_2025}.
\par
\begin{figure}[htbp]
    \centering
    \includegraphics[width=\linewidth]{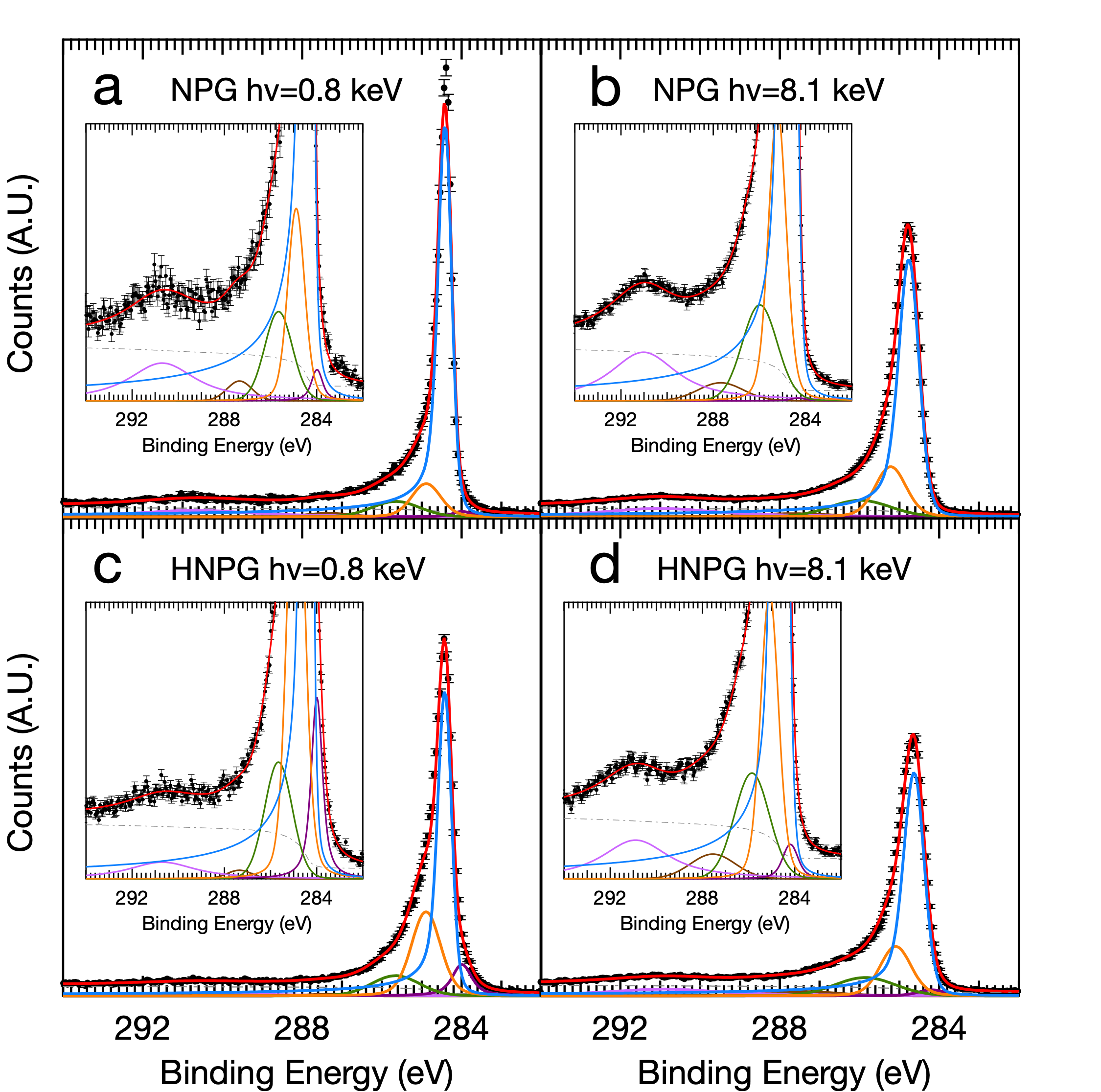}
    \caption{Single global fit of C 1s spectra for NPG (a,b) and HNPG (c,d) with zoomed insights. At 0.8 keV photon energy (a,c) and at 8.1 keV photon energy (b,d). The spectra are deconvoluted with the following components: sp² (blue), sp³ (orange), defects (purple), $\pi$ plasmon (violet), C-O (green), O-C-O (brown).}
    \label{fig:3}
\end{figure}
\begin{figure}[htbp]
    \centering
    \includegraphics[width=\linewidth]{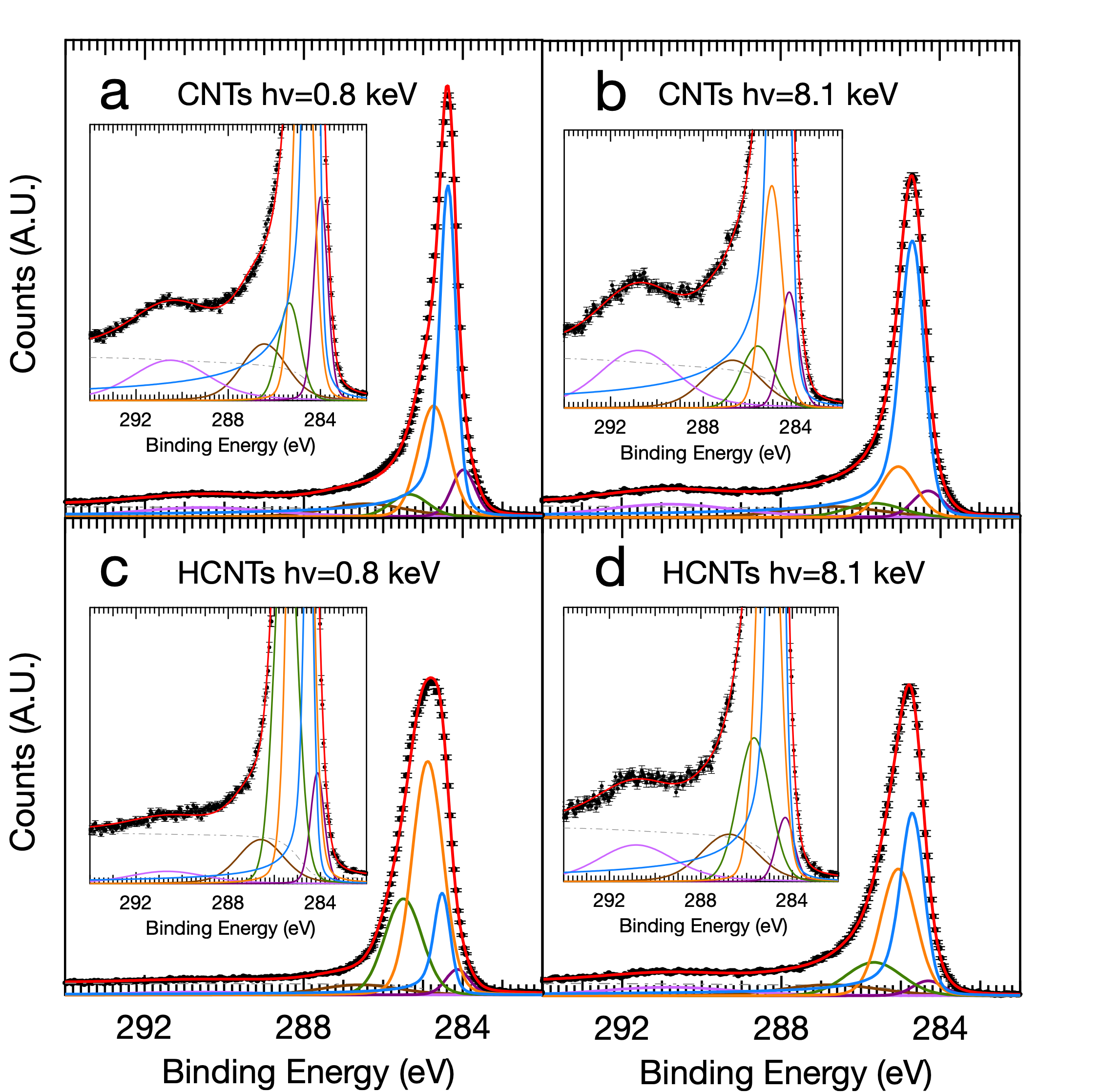}
    \caption{Single global fit of C 1s spectra for CNTs (a,b) and HCNTs (c,d) with zoomed insights. At 0.8 keV photon energy (a,c) and at 8.1 keV photon energy (b,d). The spectra are deconvoluted with the following components: sp² (blue), sp³ (orange), defects (purple), $\pi$ plasmon (violet), C–O (green), O–C–O (brown).}
    \label{fig:4}
\end{figure}

The fitting analysis performed on the C 1s spectra to deconvolve the contributions from distinct chemical environments and to quantify the evolution of sp³ hybridization upon hydrogenation is summarized in Figures~\ref{fig:3} (NPG/HNPG) and \ref{fig:4} (CNTs/HCNTs) for two selected photon energies. The fit includes the following contributions:
(i) an sp² component characteristic of planar graphene-like carbon;
(ii) an sp³ component associated with tetrahedral hybridization, arising from C–H bonding as well as from intrinsic curvature, ripples, and defects, particularly relevant in CNTs;
(iii) C–O and O–C–O components accounting for oxygen-containing species;
(iv) a defect-related component associated with dangling bonds or carbon atoms adjacent to hydrogenated sites \cite{betti_gap_2022,lizzit_dual-route_2019}; and
(v) a $\pi$-plasmon loss feature characteristic of graphitic sp² systems \cite{di_filippo_evolution_2020}. Each component is identified by its binding energy shift $\Delta E = \mathrm{BE} - \mathrm{BE}$($\mathrm{sp}^2$) relative to the sp² peak. The extracted $\Delta E$ values for NPG/HNPG and CNTs/HCNTs are summarized in Table~\ref{tab:BEshifts}, together with representative literature values.
\begin{table}[htbp]
\centering
\begin{tabular}{lccc}
\hline
%\\[-3ex]
\textbf{Comp.}
\\[-3ex]
& \shortstack{\textbf{(H)NPG}\\\textbf{(eV)}}
& \shortstack{\textbf{(H)CNTs}\\\textbf{(eV)}}
& \shortstack{\textbf{Ref.}\\\textbf{(eV)}}\\
\hline
Defects       
& $-0.41 \pm 0.03$ 
& $-0.36 \pm 0.05$ 
& $<0$ \cite{tayyab_atomic_2025,dacunto_channelling_2018} \\

sp$^3$        
& $0.49 \pm 0.03$ 
& $0.39 \pm 0.02$ 
& $0.6$--$0.8$ \cite{betti_dielectric_2023,nikitin_hydrogenation_2005,tayyab_atomic_2025,apponi_highly_2025} \\

C--O          
& $1.25 \pm 0.08$ 
& $1.0 \pm 0.08$ 
& $1.0$--$2.1$ \cite{di_filippo_evolution_2020,tayyab_atomic_2025,barinov_imaging_2009} \\

O--C--O       
& $2.95 \pm 0.05$ 
& $2.1 \pm 0.1$ 
& $2.1$--$3.8$ \cite{di_filippo_evolution_2020,tayyab_atomic_2025,barinov_imaging_2009} \\

$\pi$ plasmon 
& $6.3 \pm 0.04$ 
& $6.16 \pm 0.07$ 
& $6.0$--$6.5$ \cite{di_filippo_evolution_2020,apponi_transmission_2024} \\
\hline
\end{tabular}
\caption{Binding energy shifts ($\Delta E = \mathrm{BE} - \mathrm{BE}(\mathrm{sp}^2)$) for the different C~1s components.
Values are in eV. Literature values are reported as typical ranges.}
\label{tab:BEshifts}
\end{table}
\par
Note that a finite contribution of the sp³ spectral feature is already present in pristine samples. This ``baseline" sp³ contribution originates from intrinsic structural features of graphene-based materials, such as curvature, lattice distortions and may also include hydrogen termination of reactive sites after UHV annealing \cite{apponi_highly_2025}. Consequently, the sp³ component extracted from hydrogenated samples contains both a baseline contribution and an additional component induced by the hydrogenation process. In the following, the effects of hydrogenation are therefore assessed by comparison with the corresponding pristine reference samples.
\par
One can see that pristine NPG is dominated by the sp² component, as expected for graphene-based structures. Minor sp³ and C–O contributions are also present and show a weak increase with probing depth, possibly reflecting cleaner near-surface conditions compared to the subsurface region after annealing. The overall intensity of the oxygen-related components remains consistent with the low oxygen concentrations (a few atomic percent \cite{oxygen_note}) independently estimated from O 1s measurements. The HNPG spectrum (Fig.\ref{fig:3}c) exhibits a dominant sp² component accompanied by a significantly enhanced sp³ contribution compared to pristine NPG (Fig.\ref{fig:3}a). With increasing photon energy and probing depth, the relative intensity of the sp³ component progressively decreases, while the sp² contribution increases. This trend indicates that the hydrogen induced sp³ hybridization in NPG is predominantly localized near the surface.
\par
Pristine CNTs (Figs. \ref{fig:4}a–b result of the fit at the two extreme photon energies) also display a non-negligible sp³ component, consistent with previous reports and commonly attributed to the intrinsic curvature and structural disorder of the graphene walls \cite{Tayyab2024}. Notably, the relative intensity of this sp³ component decreases as the photon energy and thus the probing depth increases, even though the higher curvature of inner CNT walls would be expected to enhance sp³ hybridization. This suggests that the sp³ contribution in pristine CNTs is largely associated with surface related effects, potentially including residual hydrogen termination. At 0.8 keV photon energy (Fig.\ref{fig:4}c), HCNTs exhibit a dominant sp³ component accompanied by weaker sp² and C–O contributions. As the photon energy increases a larger sample depth is probed and the relative sp³ intensity decreases, indicating a reduced degree of hydrogenation in deeper regions. Nevertheless, even at the highest photon energy of 8.1 keV (Fig.\ref{fig:4}d), the sp³ fraction in HCNTs remains significantly higher than in pristine CNTs measured under the same conditions. The C–O component is most pronounced at low photon energy for both CNTs and HCNTs samples and rapidly decreases with increasing probing depth, consistent with surface-localized oxygen contamination induced by the plasma hydrogenation process \cite{whitener_review_2018}. Also in this case their contribution remains limited, in agreement with the low oxygen concentration (a few atomic percent \cite{oxygen_note}) estimated from O 1s measurements. Finally, the $\pi$-plasmon loss feature, associated with collective $\pi$-electron excitations and indicative of planar sp² order, is systematically more intense in pristine samples than in their hydrogenated counterparts, in both NPG and CNTs. Its slight increase with photon energy, even in pristine spectra, reflects the enhanced contribution of extrinsic loss features at larger probing depths, consistent with previous observations in graphite \cite{guzzo_multiple_2014}.
\begin{figure}[htbp]
    \centering    \includegraphics[width=1\linewidth]{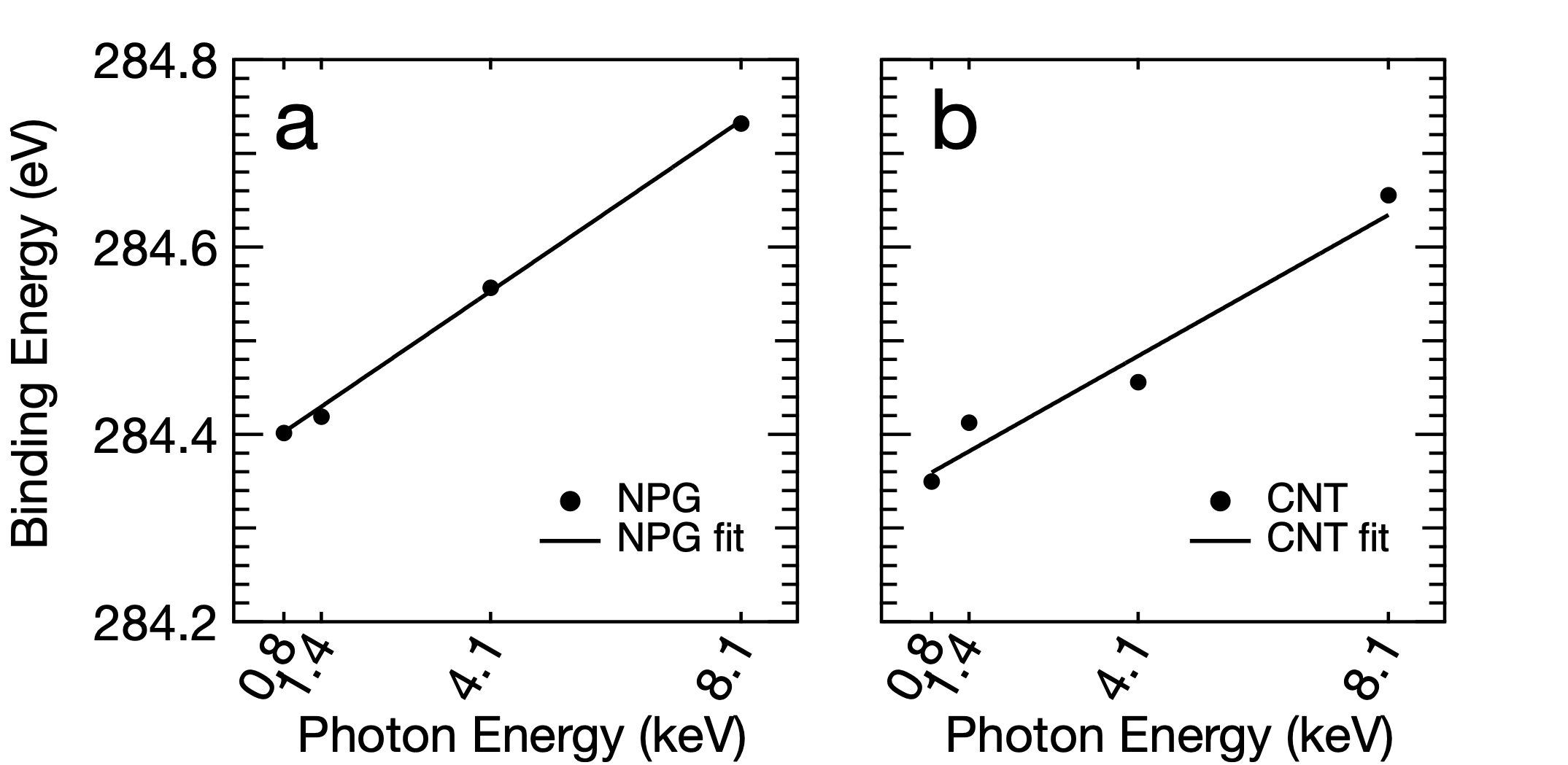}
    \caption{Photon-energy dependence of the sp² C 1s binding energy in pristine samples. Panel (a) shows NPG global-fit results and panel (b) CNT results. Solid black lines are linear fits to the data. Uncertainties are smaller than the marker size.}
    \label{fig:5}
\end{figure}
\par
It has to be noted that, at high photon energies, photoemission recoil effects become increasingly relevant and modify the observed core-level lineshape. A rigorous description of these effects can be obtained within the quantum-mechanical framework developed by Takata \cite{takata_recoil_2007} and subsequently extended to graphene by Ritarossi \cite{ritarossi}, in which the recoil spectrum is calculated by explicitly accounting for the phonon modes of the crystal. In the present work, however, we adopt a simplified description based on the semiclassical treatment proposed by Fujikawa \cite{fujikawa_2006}. Within this approach, recoil effects give rise to a rigid shift of the photoemission peak toward higher binding energy and an additional Gaussian broadening. These correspond to the first and second moments, respectively, of the recoil spectrum obtained within the full phonon-dependent model \cite{fujikawa_2006}. Although this approximation does not capture the subtle asymmetries arising from material-dependent phonon modes, it has been shown to reproduce the main spectral features of the complete recoil model remarkably well in graphite \cite{takata_recoil_2007}. Moreover, the same phenomenological approach has recently been employed to describe recoil effects in crystalline silicon and silicon carbide as well \cite{roth_recoil_2025}.
\par
According to the semiclassical treatment, the average shift of the C 1s peak $E_R$ is expected to scale linearly with the photoelectron kinetic energy $E_k$, following the free-atom recoil estimate $E_R = \frac{m}{M} E_k$, where $m/M$ is the ratio between the electron mass and the carbon atomic mass. In the present work, the binding energy separations between the different spectral components are kept fixed, while the sp² binding energy is allowed to vary for each spectrum to account for the photon-energy-dependent recoil shift. The binding energy of the sp² component in pristine NPG and CNTs spectra increases systematically with photon energy, as reported in Figs. \ref{fig:5}a–b. A linear fit of the pristine data, weighted by the uncertainties, yields slopes of ($4.6 \pm 0.1)\times10^{-5}$ for NPG and ($3.8 \pm 0.7)\times10^{-5}$ for CNTs, in good agreement with the free-carbon recoil expectation $m/M \approx 4.5\times10^{-5}$. This agreement indicates that recoil provides the dominant contribution to the observed photon-energy-dependent binding energy shift in pristine samples and that it is consistently captured within the adopted fitting approach.
\par
The systematic decrease of the sp³ spectral weight with increasing probing depth observed for both HNPG and HCNTs indicates that hydrogen induced hybridization is strongest near the surface and progressively weaker at larger depths, pointing to a depth dependent hydrogen distribution. To quantify this behavior, we employ a slightly modified version of the standard thin overlayer on substrate model commonly used in photoemission analysis \cite{Fadley1978}. In this framework, the depth profile of the sp³ is described by expressing the fraction of sp³ hybridized carbon atoms, $n(z)$, as a function of the depth coordinate $z$, defined along the surface normal and coincident with the photoelectron emission direction ($z = 0$ at the surface and increasing toward the bulk). The sp³ fraction is written as the sum of two terms: a homogeneous contribution $A$, accounting for intrinsic sp³ structural features of graphene-based materials as well as for any hydrogen induced sp³ contribution that is spatially uniform, and an additional surface-confined contribution associated with hydrogenation of a layer of finite thickness $d$. The purpose of this model is therefore not to assume the absence of bulk hydrogenation, but rather to separate homogeneous and surface-localized contributions. Under these assumptions, the depth-dependent sp³ fraction can be written as:
\begin{equation}
n(z) = A + (1-A)\, f(z)
\end{equation}
where
\begin{equation}
f(z)=
\begin{cases}
1 & \text{for}\quad 0 \le z \le d,\\
0 & \text{for}\quad z > d .
\end{cases}
\end{equation}
\par
The measured photoemission intensity is obtained by integrating the local signal over depth, weighted by the exponential attenuation of the photoelectrons along their path inside the material. Neglecting elastic scattering and in normal emission, the ratio between the sp³ intensity and the total intensity of the C 1s peak is given by
\begin{equation}
\frac{I_{\mathrm{sp}^{3}}}{I_{\mathrm{tot}}}
= \frac{\int_0^{\infty} n(z)\, e^{-z/\lambda}\, dz}{\int_0^{\infty} e^{-z/\lambda}\, dz}
= 1 + (A-1)\, e^{-d/\lambda}
\label{eqn:3}
\end{equation}
where $\lambda$ is the inelastic mean free path (IMFP) of the photoelectrons. In the limit $\lambda \gg d$, corresponding to large probing depths, the measured sp³ fraction approaches $A$, while for $\lambda \ll d$ it tends toward unity, as expected when the photoemission signal is dominated by a fully hydrogenated surface region. Rather than referencing the data directly to the photon energy, the depth dependence of the sp³ fraction is therefore expressed as a function of the corresponding IMFP, providing a physically meaningful length scale for comparison across different excitation energies. Although CNTs and NPG possess complex morphologies, the IMFP dependence on photon energies for single-wall CNTs \cite{kyriakou_electron_2009} is similar to the one of graphite. Variations in wall number and packing introduce uncertainties, but these remain within the same order of magnitude as the graphite values. On the other hand, for NPG, the measured density \cite{ito_highquality_2014} is significantly lower than the one of graphite. Nevertheless, to use graphite IMFP values and to reason in terms of an equivalent number of graphene layers may provide a reasonable approximation.
\par
Fully hydrogenated multilayer graphene shows a $\sim50 \%$ volume increase and an $\sim8 \%$ increase in molar mass relative to graphite \cite{antonov_multilayer_2016}, \cite{yartys_reversible_2025}. Plugged into the TPP-2M formula \cite{tanuma_calculation_2003}, the resulting IMFP remains essentially unchanged compared to graphite. For these reasons, we adopt the IMFP calculated on graphite and apply it to both hydrogenated and pristine layers in NPG and CNTs. Table \ref{tab:penetration} reports IMFP values for graphite. Values are linearly interpolated from a combination of IMFP calculation from optical data by Tanuma et al. \cite{tanuma_calculation_2003} at  low energies and HAXPES measurements by Kuntz et al. \cite{kunz_relative_2009} at  $h\nu$ = 8 keV.
\begin{table}
\centering
\begin{tabular}{ccc}
\hline\hline
$h\nu$ (keV) & $\text{E}_{k}$ (keV) & $\lambda$ (nm)\\
\hline
0.8  & 0.5 & 1.1 \\
1.4 & 1.1 & 2.0 \\
4.1 & 3.8 & 5.0 \\
8.1 & 7.8 & 9.2 \\
\hline\hline
\end{tabular}
\caption{IMFP of graphite ($\lambda$) for electrons with kinetic energy $\text{E}_{k}$ correspondent to C 1s photoelectron with different exciting energies $h\nu$. Values are obtained from a linear interpolation of data by Tanuma et al. \cite{tanuma_calculation_2003} at  low energies and Kuntz et al. \cite{kunz_relative_2009} at  $h\nu$ = 8 keV.}
\label{tab:penetration}
\end{table}
\begin{figure}
    \centering
    \includegraphics[width=1\linewidth]{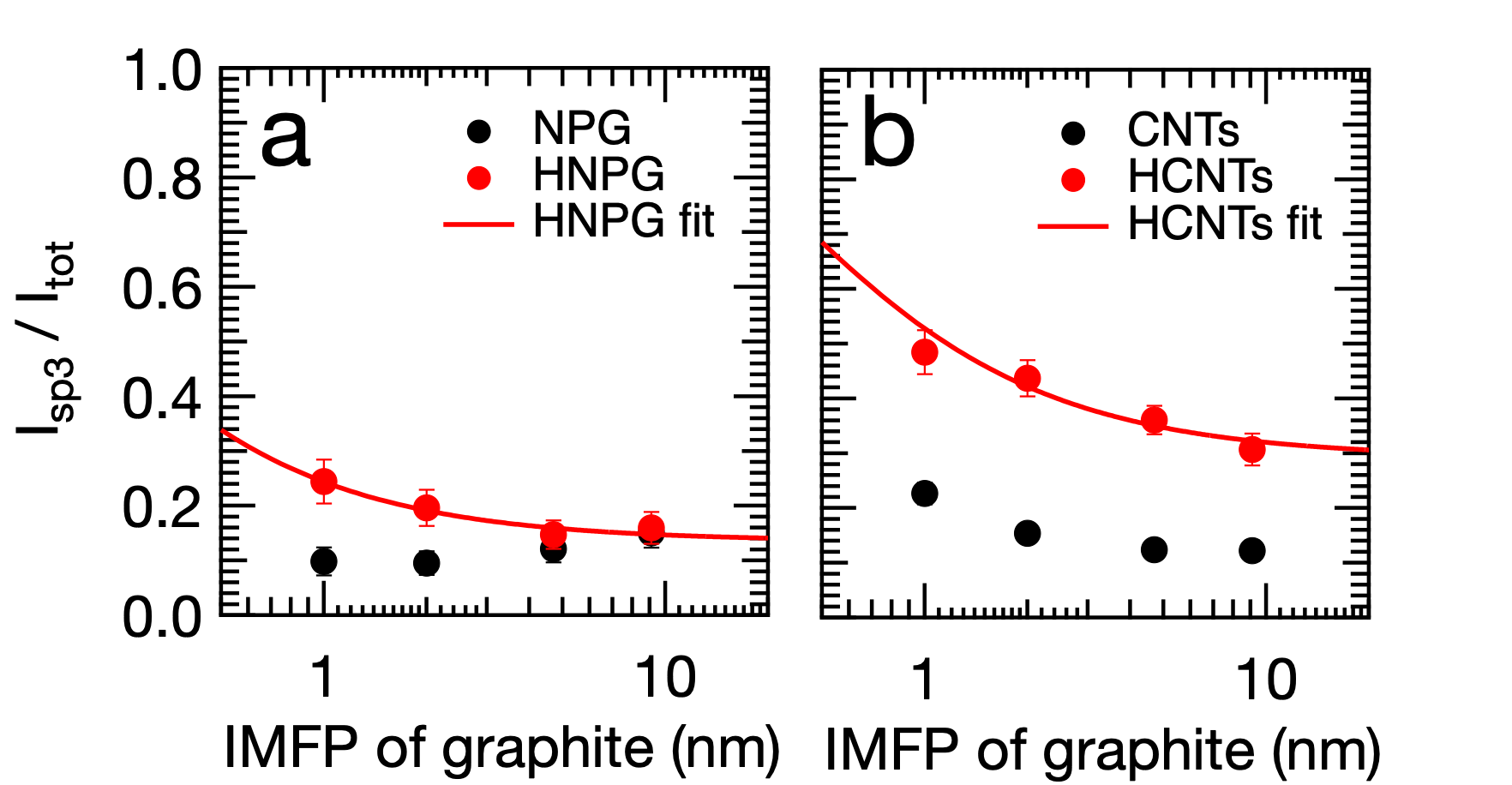}
        \caption{sp³ relative intensity with respect to the total C 1s peak area for NPG/HNPG (a) and CNTs/HCNTs (b) as a function of the IMFP of graphite. NPG (CNTs) are black errorbars, HNPG (HCNTs) are red errorbars. HNPG and HCNTs data are fitted with the function of eqn.\ref{eqn:3} (red solid line)}
    \label{fig:6}
\end{figure}
\par
In Figure \ref{fig:6}, the relative sp³ intensities for NPG/HNPG (a) and CNTs/HCNTs (b) are reported as a function of the IMFP of graphite. The data for the hydrogenated samples are fitted using the model described by Eq. \ref{eqn:3}. For HNPG, the fit yields an effective hydrogenated overlayer thickness of $d = 0.13 \pm 0.06$ nm, corresponding to a fraction of a graphene layer in graphite ($\sim0.34$ nm \cite{baskin_lattice_1955}). The homogeneous sp³ contribution is $A = 0.14 \pm 0.02$, consistent with the sp³ concentration observed in pristine NPG, approximately constant vs IMFP (Fig. \ref{fig:6}a). This indicates that hydrogenation introduces a surface localized sp³ component, while the sp³ fraction remains essentially unchanged in the inner layers. This result is consistent with hydrogen adsorption being confined to the outermost graphene surface, with no significant evidence of penetration into the internal volume of the nanoporous network.
\par
For HCNTs, the sp³ fraction is systematically higher than in pristine CNTs at all probing depths. The fitted overlayer thickness is d = $0.4 \pm 0.1$ nm, corresponding to approximately one graphene layer, while the homogeneous sp³ contribution amounts to $A = 0.29 \pm 0.02$, substantially larger than the $\sim 0.1$ relative sp³ intensity observed in pristine CNTs. Interpreting these values is more complex due to the morphology of vertically aligned multi-walled CNTs, which include multiple concentric graphene shells and inter-wall spacing. Nevertheless, the results are consistent with a strong hydrogenation of the outermost CNTs walls. At the same time, the relatively small value of $d$ suggests that hydrogen penetration into the innermost CNT walls or along the tube's axis is limited. This interpretation does not contradict earlier experimental and theoretical studies reporting uniform hydrogenation in single-wall CNTs, where hydrogen readily accesses the entire carbon network \cite{nikitin_hydrogenation_2005,nikitin_hydrogen_2008}. Rather, the present results highlight a key difference between single and multi-walled systems: in multi-walled CNTs, the inner walls appear significantly less accessible to hydrogen under plasma exposure, leading to a non-uniform depth profile. It is also important to distinguish the present hydrogenation pathway from chemically aggressive processes such as Birch reduction, where hydrogenation of inner CNT walls has been observed by transmission electron microscopy \cite{pekker_hydrogenation_2001}. In such cases, hydrogen adsorption occurs under strongly reducing conditions, which are fundamentally different from the plasma hydrogenation employed here.
\par
More broadly, these findings relate to ongoing discussions on the permeability of graphene-based materials to hydrogen. While monolayer graphene has been reported to be permeable to molecular hydrogen at high pressure \cite{sun_limits_2020}, our observations are consistent with limited permeability of graphene to atomic hydrogen, leading to preferential hydrogenation of the outermost layer. For NPG, this effect may be further enhanced by the reduced accessibility of the internal porous network; its bicontinuous monolithic architecture \cite{ito_highquality_2014} can hinder the diffusion of reactive hydrogen species toward the internal surfaces.
\section{Conclusions}
The present results suggest that the hydrogenation of both nanoporous graphene and multi-walled carbon nanotubes leads to the formation of a sp³-rich region which is not homogeneous in depth. In HNPG, the effective hydrogenated thickness corresponds to a fraction of a graphene layer, indicating that C–H bond formation is confined to the outermost accessible surface, with no clear evidence for penetration into the internal graphene network.
\par
In HCNTs, the hydrogenated region extends to a thickness comparable to approximately one graphene layer, and the overall sp³ fraction is systematically higher than in pristine CNTs even at the largest probing depths. These findings are consistent with efficient hydrogenation of the outer CNTs walls and with a partial contribution from regions beyond the immediate surface, such as inter-wall spaces accessible to atomic hydrogen. 
\par
Notably, HCNTs show higher overall hydrogen content than HNPG. This difference may reflect the larger accessible surface area of vertically aligned CNT bundles; however, a direct quantitative comparison is complicated by the different hydrogenation procedures employed for the two materials (plasma exposure for CNTs and atomic hydrogen cracking for NPG). Notably, despite differences in material morphology and hydrogenation routes, both systems exhibit predominantly surface-localized hydrogen adsorption, with penetration depths significantly smaller than the escape depth of keV photoelectrons. While the limited hydrogenation depth reduces the extent of bulk functionalization, the two architectures offer complementary advantages: the bicontinuous graphene network of NPG can help preserve electrical conductivity while enabling surface functionalization, whereas the large accessible surface area of vertically aligned CNT bundles provides a high density of chemically active sites, making them attractive for hydrogen-storage applications.
\section*{Acknowledgments}
We acknowledge financial support under the National
Recovery and Resilience Plan (NRRP), Mission 4, Component 2, Investment 1.1, Call for tender No. 104 published on 2 February 2022 by the Italian Ministry of
University and Research (MUR), funded by the European Union – NextGenerationEU – Project Title PACE
(20227F53E4), CUP F53C24000790006, Grant Assignment Decree No. 20429 adopted on 6 November 2024 by
the Italian Ministry of University and Research (MUR).
This work was partially supported by the Italian Ministry of University and Research (MUR) under the Grant
of Excellence Departments, Art. 1, commi 314–337, Law
232/2016, to the Department of Science, Roma Tre University. The support from Diamond Light Source, beamline I09 is gratefully acknowledged. This work was partially supported byJSPS-Kakenhi (JP23K17661, JP24H00478).
 \section*{Author Contributions}
 O.C.: Investigation, Methodology, Data curation, Formal analysis, Visualization, Writing -- original draft. A.A.: Conceptualization, Investigation, Formal analysis, Methodology, Writing -- review and editing. L.C.: Methodology, Resources, Writing -- review and editing. D.P.: Investigation, Data curation, Formal analysis, Writing -- review and editing. S.R.: Formal analysis, Methodology, Writing -- review and editing. F.P.: Resources, Writing -- review and editing. I.R.: Resources, Writing -- review and editing. T.-L.L.: Resources, Writing -- review and editing. C.M.: Conceptualization, Resources, Writing -- review and editing. G.C.: Resources, Writing -- review and editing. F.O.: Conceptualization, Investigation, Data curation, Methodology, Supervision, Writing -- review and editing. A.R.: Conceptualization, Investigation, Supervision, Project administration, Funding acquisition, Writing -- review and editing.
\section*{Data Availability}
The data used in this work are available from the corresponding author upon reasonable request.
\bibliographystyle{apsrev4-2}
\bibliography{references}

@incollection{Fadley1978,
  author       = {Fadley, C. S.},
  title        = {X-ray Photoelectron Spectroscopy: Progress and Perspectives},
  booktitle    = {Electron Spectroscopy: Theory, Techniques and Applications},
  editor       = {Brundle, C. R. and Baker, A. D.},
  publisher    = {Academic Press},
  address      = {London},
  volume       = {2},
  year         = {1978},
  pages        = {1--156}
}

@article{Tayyab2024,
  author       = {Tayyab, Sammar and Apponi, Alice and Betti, Maria Grazia and Blundo, Elena and Cavoto, Gianluca and Frisenda, Riccardo and Jim\'enez-Ar\'evalo, Nuria and Mariani, Carlo and Pandolfi, Francesco and Polimeni, Antonio and Rago, Ilaria and Ruocco, Alessandro and Sbroscia, Marco and Yadav, Ravi Prakash},
  title        = {Spectromicroscopy Study of Induced Defects in Ion-Bombarded Highly Aligned Carbon Nanotubes},
  journal      = {Nanomaterials},
  year         = {2024},
  volume       = {14},
  number       = {1},
  pages        = {77},
  doi          = {10.3390/nano14010077},
  url          = {https://www.mdpi.com/2079-4991/14/1/77}
}

@misc{arias_darkmatter_2026,
author        = {Tom'as Arias and Antonino Bellinvia and Gianluca Cavoto and Angelo Esposito and Francesco Pandolfi and Guglielmo Papiri and Antonio D. Polosa and Tyler Wu},
title         = {Hydrogenated Carbon Structures as Directional Sub-GeV Dark Matter Detectors},
year          = {2026},
eprint        = {2602.02694},
archivePrefix = {arXiv},
primaryClass  = {hep-ph}
}

@article{lizzit_dual-route_2019,
	title = {Dual-{Route} {Hydrogenation} of the {Graphene}/{Ni} {Interface}},
	issn = {1936-0851, 1936-086X},
	url = {https://pubs.acs.org/doi/10.1021/acsnano.8b07996},
	doi = {10.1021/acsnano.8b07996},
	language = {en},
	urldate = {2023-02-14},
	journal = {ACS Nano},
	author = {Lizzit, Daniel and Trioni, Mario I. and Bignardi, Luca and Lacovig, Paolo and Lizzit, Silvano and Martinazzo, Rocco and Larciprete, Rosanna},
	month = jan,
	year = {2019},
	pages = {acsnano.8b07996},
}

@article{sofo_graphane_2007,
	title = {Graphane: {A} two-dimensional hydrocarbon},
	volume = {75},
	copyright = {http://link.aps.org/licenses/aps-default-license},
	issn = {1098-0121, 1550-235X},
	shorttitle = {Graphane},
	url = {https://link.aps.org/doi/10.1103/PhysRevB.75.153401},
	doi = {10.1103/PhysRevB.75.153401},
	language = {en},
	number = {15},
	urldate = {2025-10-29},
	journal = {Phys. Rev. B},
	author = {Sofo, Jorge O. and Chaudhari, Ajay S. and Barber, Greg D.},
	month = apr,
	year = {2007},
	pages = {153401},
}

@article{elias_control_2009,
	title = {Control of {Graphene}'s {Properties} by {Reversible} {Hydrogenation}: {Evidence} for {Graphane}},
	volume = {323},
	issn = {0036-8075, 1095-9203},
	shorttitle = {Control of {Graphene}'s {Properties} by {Reversible} {Hydrogenation}},
	url = {https://www.science.org/doi/10.1126/science.1167130},
	doi = {10.1126/science.1167130},
	language = {en},
	number = {5914},
	urldate = {2025-10-29},
	journal = {Science},
	author = {Elias, D. C. and Nair, R. R. and Mohiuddin, T. M. G. and Morozov, S. V. and Blake, P. and Halsall, M. P. and Ferrari, A. C. and Boukhvalov, D. W. and Katsnelson, M. I. and Geim, A. K. and Novoselov, K. S.},
	month = jan,
	year = {2009},
	pages = {610--613},
}

@article{nikitin_hydrogenation_2005,
	title = {Hydrogenation of {Single}-{Walled} {Carbon} {Nanotubes}},
	volume = {95},
	copyright = {http://link.aps.org/licenses/aps-default-license},
	issn = {0031-9007, 1079-7114},
	url = {https://link.aps.org/doi/10.1103/PhysRevLett.95.225507},
	doi = {10.1103/PhysRevLett.95.225507},
	language = {en},
	number = {22},
	urldate = {2025-10-29},
	journal = {Phys. Rev. Lett.},
	author = {Nikitin, A. and Ogasawara, H. and Mann, D. and Denecke, R. and Zhang, Z. and Dai, H. and Cho, K. and Nilsson, A.},
	month = nov,
	year = {2005},
	pages = {225507},
}

@article{tayyab_atomic_2025,
	title = {Atomic deuterium bonding to multi-walled carbon nano tubes},
	volume = {162},
	issn = {0021-9606, 1089-7690},
	url = {https://pubs.aip.org/jcp/article/162/19/194704/3346911/Atomic-deuterium-bonding-to-multi-walled-carbon},
	doi = {10.1063/5.0250642},
	language = {en},
	number = {19},
	urldate = {2025-10-29},
	journal = {The Journal of Chemical Physics},
	author = {Tayyab, Sammar and Apponi, Alice and Betti, Maria Grazia and Blundo, Elena and Castellano, Orlando and Cavoto, Gianluca and Pandolfi, Francesco and Polimeni, Antonio and Rago, Ilaria and Ruocco, Alessandro and Yadav, Ravi Prakash and Mariani, Carlo},
	month = may,
	year = {2025},
	pages = {194704},
}

@article{betti_gap_2022,
	title = {Gap {Opening} in {Double}-{Sided} {Highly} {Hydrogenated} {Free}-{Standing} {Graphene}},
	volume = {22},
	copyright = {https://creativecommons.org/licenses/by/4.0/},
	issn = {1530-6984, 1530-6992},
	url = {https://pubs.acs.org/doi/10.1021/acs.nanolett.2c00162},
	doi = {10.1021/acs.nanolett.2c00162},
	language = {en},
	number = {7},
	urldate = {2025-10-29},
	journal = {Nano Lett.},
	author = {Betti, Maria Grazia and Placidi, Ernesto and Izzo, Chiara and Blundo, Elena and Polimeni, Antonio and Sbroscia, Marco and Avila, Jos{\'e} and Dudin, Pavel and Hu, Kailong and Ito, Yoshikazu and Prezzi, Deborah and Bonacci, Miki and Molinari, Elisa and Mariani, Carlo},
	month = apr,
	year = {2022},
	pages = {2971--2977},
}

@article{betti_homogeneous_2022,
	title = {Homogeneous {Spatial} {Distribution} of {Deuterium} {Chemisorbed} on {Free}-{Standing} {Graphene}},
	volume = {12},
	issn = {2079-4991},
	url = {https://www.mdpi.com/2079-4991/12/15/2613},
	doi = {10.3390/nano12152613},
	language = {en},
	number = {15},
	urldate = {2025-10-29},
	journal = {Nanomaterials},
	author = {Betti, Maria Grazia and Blundo, Elena and De Luca, Marta and Felici, Marco and Frisenda, Riccardo and Ito, Yoshikazu and Jeong, Samuel and Marchiani, Dario and Mariani, Carlo and Polimeni, Antonio and Sbroscia, Marco and Trequattrini, Francesco and Trotta, Rinaldo},
	month = jul,
	year = {2022},
	pages = {2613},
}

@article{betti_neutrino_2019,
	title = {Neutrino physics with the {PTOLEMY} project: active neutrino properties and the light sterile case},
	volume = {2019},
	copyright = {http://iopscience.iop.org/info/page/text-and-data-mining},
	issn = {1475-7516},
	shorttitle = {Neutrino physics with the {PTOLEMY} project},
	url = {https://iopscience.iop.org/article/10.1088/1475-7516/2019/07/047},
	doi = {10.1088/1475-7516/2019/07/047},
	number = {07},
	urldate = {2025-10-29},
	journal = {J. Cosmol. Astropart. Phys.},
	author = {Betti, M.G. and Biasotti, M. and Bosc{\'a}, A. and Calle, F. and Canci, N. and Cavoto, G. and Chang, C. and Cocco, A.G. and Colijn, A.P. and Conrad, J. and D'Ambrosio, N. and Groot, N. De and De Salas, P.F. and Faverzani, M. and Ferella, A. and Ferri, E. and Garcia-Abia, P. and Garc{\'i}a-Cort{\'e}s, I. and Gomez-Tejedor, G. Garcia and Gariazzo, S. and Gatti, F. and Gentile, C. and Giachero, A. and Gudmundsson, J.E. and Hochberg, Y. and Kahn, Y. and Kievsky, A. and Lisanti, M. and Mancini-Terracciano, C. and Mangano, G. and Marcucci, L.E. and Mariani, C. and Mart{\'i}nez, J. and Messina, M. and Molinero-Vela, A. and Monticone, E. and Moro{\~n}o, A. and Nucciotti, A. and Pandolfi, F. and Parlati, S. and Pastor, S. and Pedr{\'o}s, J. and Heros, C. P{\'e}rez De Los and Pisanti, O. and Polosa, A.D. and Puiu, A. and Rago, I. and Raitses, Y. and Rajteri, M. and Rossi, N. and Rucandio, I. and Santorelli, R. and Schaeffner, K. and Tully, C.G. and Viviani, M. and Zhao, F. and Zurek, K.M.},
	month = jul,
	year = {2019},
	pages = {047--047},
}

@article{ito_highquality_2014,
	title = {High-{Quality} {Three}-{Dimensional} {Nanoporous} {Graphene}},
	volume = {53},
	copyright = {http://onlinelibrary.wiley.com/termsAndConditions\#vor},
	issn = {1433-7851, 1521-3773},
	url = {https://onlinelibrary.wiley.com/doi/10.1002/anie.201402662},
	doi = {10.1002/anie.201402662},
	language = {en},
	number = {19},
	urldate = {2025-10-29},
	journal = {Angew Chem Int Ed},
	author = {Ito, Yoshikazu and Tanabe, Yoichi and Qiu, H.-J. and Sugawara, Katsuaki and Heguri, Satoshi and Tu, Ngoc Han and Huynh, Khuong Kim and Fujita, Takeshi and Takahashi, Takashi and Tanigaki, Katsumi and Chen, Mingwei},
	month = may,
	year = {2014},
	pages = {4822--4826},
}

@article{schifano_plasma-etched_2023,
	title = {Plasma-{Etched} {Vertically} {Aligned} {CNTs} with {Enhanced} {Antibacterial} {Power}},
	volume = {13},
	issn = {2079-4991},
	url = {https://www.mdpi.com/2079-4991/13/6/1081},
	doi = {10.3390/nano13061081},
	language = {en},
	number = {6},
	urldate = {2025-10-29},
	journal = {Nanomaterials},
	author = {Schifano, Emily and Cavoto, Gianluca and Pandolfi, Francesco and Pettinari, Giorgio and Apponi, Alice and Ruocco, Alessandro and Uccelletti, Daniela and Rago, Ilaria},
	month = mar,
	year = {2023},
	pages = {1081},
}

@article{apponi_implementation_2022,
	title = {Implementation and optimization of the {PTOLEMY} transverse drift electromagnetic filter},
	volume = {17},
	issn = {1748-0221},
	url = {https://iopscience.iop.org/article/10.1088/1748-0221/17/05/P05021},
	doi = {10.1088/1748-0221/17/05/P05021},
	number = {05},
	urldate = {2025-10-29},
	journal = {J. Inst.},
	author = {Apponi, A. and Betti, M.G. and Borghesi, M. and Canci, N. and Cavoto, G. and Chang, C. and Chung, W. and Cocco, A.G. and Colijn, A.P. and D'Ambrosio, N. and De Groot, N. and Faverzani, M. and Ferella, A. and Ferri, E. and Ficcadenti, L. and Gariazzo, S. and Gatti, F. and Gentile, C. and Giachero, A. and Hochberg, Y. and Kahn, Y. and Kievsky, A. and Lisanti, M. and Mangano, G. and Marcucci, L.E. and Mariani, C. and Messina, M. and Monticone, E. and Nucciotti, A. and Orlandi, D. and Pandolfi, F. and Parlati, S. and P{\'e}rez De Los Heros, C. and Pisanti, O. and Polosa, A.D. and Puiu, A. and Rago, I. and Raitses, Y. and Rajteri, M. and Rossi, N. and Rozwadowska, K. and Ruocco, A. and Strid, C.F. and Tan, A. and Tully, C.G. and Viviani, M. and Zeitler, U. and Zhao, F.},
	month = may,
	year = {2022},
	pages = {P05021},
}

@article{takata_recoil_2007,
	title = {Recoil effects of photoelectrons in a solid},
	volume = {75},
	copyright = {http://link.aps.org/licenses/aps-default-license},
	issn = {1098-0121, 1550-235X},
	url = {https://link.aps.org/doi/10.1103/PhysRevB.75.233404},
	doi = {10.1103/PhysRevB.75.233404},
	language = {en},
	number = {23},
	urldate = {2026-01-02},
	journal = {Phys. Rev. B},
	author = {Takata, Y. and Kayanuma, Y. and Yabashi, M. and Tamasaku, K. and Nishino, Y. and Miwa, D. and Harada, Y. and Horiba, K. and Shin, S. and Tanaka, S. and Ikenaga, E. and Kobayashi, K. and Senba, Y. and Ohashi, H. and Ishikawa, T.},
	month = jun,
	year = {2007},
	pages = {233404},
}

@article{fujikawa_2006,

	title = {Theory of recoil effects of elastically scattered electrons and of photoelectrons},
	volume = {151},
	number = {3},
	issn = {0368-2048},
	url = {https://www.sciencedirect.com/science/article/pii/S0368204805005232},
	doi = {10.1016/j.elspec.2005.11.011},
	language = {en},
	journal = {Journal of Electron Spectroscopy and Related Phenomena},
	author = {Fujikawa, Takashi and Suzuki, Rie and K{\"o}v{\'e}r, L.},
	year = {2006},
	pages = {170--177},

}

@article{roth_recoil_2025,

	title = {Recoil effects in high energy photoemission of solids − Revisited},
	volume = {281},
	issn = {0368-2048},
	url = {https://www.sciencedirect.com/science/article/pii/S0368204825000313},
	doi = {10.1016/j.elspec.2025.147544},
	language = {en},
	journal = {Journal of Electron Spectroscopy and Related Phenomena},
	author = {Roth, F. and Potorochin, D. and Gloskovskii, A. and Schlueter, C. and Wenthaus, L. and Molodtsov, S. and Drube, W. and Eberhardt, W.},
	year = {2025},
	pages = {147544},

}

@article{whitener_review_2018,
	title = {Review {Article}: {Hydrogenated} graphene: {A} user{\textquoteright}s guide},
	volume = {36},
	issn = {0734-2101, 1520-8559},
	shorttitle = {Review {Article}},
	url = {https://pubs.aip.org/jva/article/36/5/05G401/278/Review-Article-Hydrogenated-graphene-A-user-s},
	doi = {10.1116/1.5034433},
	language = {en},
	number = {5},
	urldate = {2026-01-02},
	journal = {Journal of Vacuum Science \& Technology A: Vacuum, Surfaces, and Films},
	author = {Whitener, Keith E.},
	month = sep,
	year = {2018},
	pages = {05G401},
}

@article{guzzo_multiple_2014,
	title = {Multiple satellites in materials with complex plasmon spectra: {From} graphite to graphene},
	volume = {89},
	copyright = {http://link.aps.org/licenses/aps-default-license},
	issn = {1098-0121, 1550-235X},
	shorttitle = {Multiple satellites in materials with complex plasmon spectra},
	url = {https://link.aps.org/doi/10.1103/PhysRevB.89.085425},
	doi = {10.1103/PhysRevB.89.085425},
	language = {en},
	number = {8},
	urldate = {2026-01-02},
	journal = {Phys. Rev. B},
	author = {Guzzo, Matteo and Kas, Joshua J. and Sponza, Lorenzo and Giorgetti, Christine and Sottile, Francesco and Pierucci, Debora and Silly, Mathieu G. and Sirotti, Fausto and Rehr, John J. and Reining, Lucia},
	month = feb,
	year = {2014},
	pages = {085425},
}

@article{kyriakou_electron_2009,
	title = {Electron inelastic mean free paths for carbon nanotubes from optical data},
	volume = {94},
	issn = {0003-6951, 1077-3118},
	url = {https://pubs.aip.org/apl/article/94/26/263113/324310/Electron-inelastic-mean-free-paths-for-carbon},
	doi = {10.1063/1.3167819},
	language = {en},
	number = {26},
	urldate = {2026-01-02},
	journal = {Applied Physics Letters},
	author = {Kyriakou, Ioanna and Emfietzoglou, Dimitris and Garcia-Molina, Rafael and Abril, Isabel and Kostarelos, Kostas},
	month = jun,
	year = {2009},
	pages = {263113},
}

@article{ito_multifunctional_2015,
	title = {Multifunctional {Porous} {Graphene} for {High}-{Efficiency} {Steam} {Generation} by {Heat} {Localization}},
	volume = {27},
	copyright = {http://onlinelibrary.wiley.com/termsAndConditions\#vor},
	issn = {0935-9648, 1521-4095},
	url = {https://advanced.onlinelibrary.wiley.com/doi/10.1002/adma.201501832},
	doi = {10.1002/adma.201501832},
	language = {en},
	number = {29},
	urldate = {2026-01-02},
	journal = {Advanced Materials},
	author = {Ito, Yoshikazu and Tanabe, Yoichi and Han, Jiuhui and Fujita, Takeshi and Tanigaki, Katsumi and Chen, Mingwei},
	month = aug,
	year = {2015},
	pages = {4302--4307},
}

@article{antonov_multilayer_2016,
	title = {Multilayer graphane synthesized under high hydrogen pressure},
	volume = {100},
	issn = {00086223},
	url = {https://linkinghub.elsevier.com/retrieve/pii/S0008622315305315},
	doi = {10.1016/j.carbon.2015.12.051},
	language = {en},
	urldate = {2026-01-02},
	journal = {Carbon},
	author = {Antonov, V.E. and Bashkin, I.O. and Bazhenov, A.V. and Bulychev, B.M. and Fedotov, V.K. and Fursova, T.N. and Kolesnikov, A.I. and Kulakov, V.I. and Lukashev, R.V. and Matveev, D.V. and Sakharov, M.K. and Shulga, Y.M.},
	month = apr,
	year = {2016},
	pages = {465--473},
}

@article{yartys_reversible_2025,
	title = {Reversible hydrogen storage in multilayer graphane: {Lattice} dynamics, compressibility, and heat capacity studies},
	volume = {332},
	issn = {02540584},
	shorttitle = {Reversible hydrogen storage in multilayer graphane},
	url = {https://linkinghub.elsevier.com/retrieve/pii/S0254058424013609},
	doi = {10.1016/j.matchemphys.2024.130232},
	language = {en},
	urldate = {2026-01-02},
	journal = {Materials Chemistry and Physics},
	author = {Yartys, Volodymyr A. and Antonov, Vladimir E. and Bulychev, Boris M. and Efimchenko, Vadim S. and Kulakov, Valery I. and Kuzovnikov, Mikhail A. and Howie, Ross T. and Shuttleworth, Hannah A. and Holin, Mylaine and Rae, Rebecca and Stone, Matthew B. and Tarasov, Boris P. and Usmanov, Radion I. and Kolesnikov, Alexander I.},
	month = feb,
	year = {2025},
	pages = {130232},
}

@article{tanuma_calculation_2003,
	title = {Calculation of electron inelastic mean free paths ({IMFPs}) {VII}. {Reliability} of the {TPP}-{2M} {IMFP} predictive equation},
	volume = {35},
	copyright = {http://onlinelibrary.wiley.com/termsAndConditions\#vor},
	issn = {0142-2421, 1096-9918},
	url = {https://analyticalsciencejournals.onlinelibrary.wiley.com/doi/10.1002/sia.1526},
	doi = {10.1002/sia.1526},
	language = {en},
	number = {3},
	urldate = {2026-01-02},
	journal = {Surface \& Interface Analysis},
	author = {Tanuma, S. and Powell, C. J. and Penn, D. R.},
	month = mar,
	year = {2003},
	pages = {268--275},
}

@article{kunz_relative_2009,
	title = {Relative electron inelastic mean free paths for diamond and graphite at {8keV} and intrinsic contributions to the energy-loss},
	volume = {173},
	copyright = {https://www.elsevier.com/tdm/userlicense/1.0/},
	issn = {03682048},
	url = {https://linkinghub.elsevier.com/retrieve/pii/S0368204809000851},
	doi = {10.1016/j.elspec.2009.03.022},
	language = {en},
	number = {1},
	urldate = {2026-01-02},
	journal = {Journal of Electron Spectroscopy and Related Phenomena},
	author = {Kunz, C. and Cowie, B.C.C. and Drube, W. and Lee, T.-L. and Thiess, S. and Wild, C. and Zegenhagen, J.},
	month = jun,
	year = {2009},
	pages = {29--39},
}

@article{lee_two-color_2018,
	title = {A {Two}-{Color} {Beamline} for {Electron} {Spectroscopies} at {Diamond} {Light} {Source}},
	volume = {31},
	issn = {0894-0886, 1931-7344},
	url = {https://www.tandfonline.com/doi/full/10.1080/08940886.2018.1483653},
	doi = {10.1080/08940886.2018.1483653},
	language = {en},
	number = {4},
	urldate = {2026-01-02},
	journal = {Synchrotron Radiation News},
	author = {Lee, Tien-Lin and Duncan, David A.},
	month = jul,
	year = {2018},
	pages = {16--22},
}

@article{cecchini_quantitative_2025,
	title = {Quantitative correlation between carbon nanotube tip morphology and field emission properties at cryogenic temperature},
	volume = {17},
	issn = {2040-3364, 2040-3372},
	url = {https://xlink.rsc.org/?DOI=D5NR02221E},
	doi = {10.1039/D5NR02221E},
	language = {en},
	number = {36},
	urldate = {2026-01-02},
	journal = {Nanoscale},
	author = {Cecchini, Luca and Pepe, Carlo and Corcione, Benedetta and Castellano, Orlando and Paoloni, Daniele and Malnati, Federico and Cavoto, Gianluca and Carminati, Marco and Fiorini, Carlo and Pettinari, Giorgio and Yadav, Ravi Prakash and Rago, Ilaria and Apponi, Alice and Puiu, Andrei and Mariani, Carlo and Rajteri, Mauro and Ruocco, Alessandro and Pandolfi, Francesco},
	year = {2025},
	pages = {21260--21267},
}

@article{apponi_stability_2026,
	title = {Stability of highly hydrogenated monolayer graphene in ultra-high vacuum and in air},
	volume = {723},
	issn = {01694332},
	url = {https://linkinghub.elsevier.com/retrieve/pii/S0169433225033756},
	doi = {10.1016/j.apsusc.2025.165658},
	language = {en},
	urldate = {2026-01-02},
	journal = {Applied Surface Science},
	author = {Apponi, Alice and Castellano, Orlando and Paoloni, Daniele and Convertino, Domenica and Mishra, Neeraj and Coletti, Camilla and Casale, Andrea and Cecchini, Luca and Cocco, Alfredo G. and Corcione, Benedetta and D{\textquoteright}Ambrosio, Nicola and Esposito, Angelo and Ferella, Alfredo and Messina, Marcello and Pandolfi, Francesco and Pofi, Francesca and Rago, Ilaria and Rossi, Nicola and Tayyab, Sammar and Yadav, Ravi Prakash and Virzi, Federico and Mariani, Carlo and Cavoto, Gianluca and Ruocco, Alessandro},
	month = mar,
	year = {2026},
	pages = {165658},
}

@misc{oxygen_note,
  author = {},
  title = {},
  year = {},
  note = {Oxygen atomic concentrations were estimated from Shirley-background-subtracted O 1s/C 1s intensity ratios, normalized to the acquisition time and corrected for photon-energy-dependent photoionization cross sections and inelastic mean free paths (IMFPs). The resulting oxygen contents are approximately 1 at.\% for pristine NPG and CNTs, 2 at.\% for HCNTs, and 6 at.\% for HNPG.}
}

@misc{apponi_highly_2025,
      title={A Wide Optical-Gap in Fully $sp^3$-Like Hydrogenated Monolayer Graphene}, 
      author={Alice Apponi and Orlando Castellano and Daniele Paoloni and Domenica Convertino and Neeraj Mishra and Camilla Coletti and Carlo Mariani and Alessandro Ruocco},
      year={2026},
      eprint={2504.10238},
      archivePrefix={arXiv},
      primaryClass={cond-mat.mtrl-sci},
      url={https://arxiv.org/abs/2504.10238}, 
}

@article{di_filippo_evolution_2020,
	title = {The evolution of hydrogen induced defects and the restoration of $\pi$ -plasmon as a monitor of the thermal reduction of graphene oxide},
	volume = {512},
	issn = {01694332},
	url = {https://linkinghub.elsevier.com/retrieve/pii/S0169433220303615},
	doi = {10.1016/j.apsusc.2020.145605},
	language = {en},
	urldate = {2026-01-02},
	journal = {Applied Surface Science},
	author = {Di Filippo, Gianluca and Liscio, Andrea and Ruocco, Alessandro},
	month = may,
	year = {2020},
	pages = {145605},
}

@article{sun_limits_2020,
	title = {Limits on gas impermeability of graphene},
	volume = {579},
	issn = {0028-0836, 1476-4687},
	url = {https://www.nature.com/articles/s41586-020-2070-x},
	doi = {10.1038/s41586-020-2070-x},
	language = {en},
	number = {7798},
	urldate = {2026-01-19},
	journal = {Nature},
	author = {Sun, P. Z. and Yang, Q. and Kuang, W. J. and Stebunov, Y. V. and Xiong, W. Q. and Yu, J. and Nair, R. R. and Katsnelson, M. I. and Yuan, S. J. and Grigorieva, I. V. and Lozada-Hidalgo, M. and Wang, F. C. and Geim, A. K.},
	month = mar,
	year = {2020},
	pages = {229--232},
}

@article{pekker_hydrogenation_2001,
	title = {Hydrogenation of {Carbon} {Nanotubes} and {Graphite} in {Liquid} {Ammonia}},
	volume = {105},
	issn = {1520-6106, 1520-5207},
	url = {https://pubs.acs.org/doi/10.1021/jp010642o},
	doi = {10.1021/jp010642o},
	language = {en},
	number = {33},
	urldate = {2026-01-19},
	journal = {J. Phys. Chem. B},
	author = {Pekker, S. and Salvetat, J.-P. and Jakab, E. and Bonard, J.-M. and Forr{\'o}, L.},
	month = aug,
	year = {2001},
	pages = {7938--7943},
}

@article{dacunto_channelling_2018,
	title = {Channelling and induced defects at ion-bombarded aligned multiwall carbon nanotubes},
	volume = {139},
	issn = {00086223},
	url = {https://linkinghub.elsevier.com/retrieve/pii/S0008622318306742},
	doi = {10.1016/j.carbon.2018.07.032},
	language = {en},
	urldate = {2026-02-06},
	journal = {Carbon},
	author = {D'Acunto, Giulio and Ripanti, Francesca and Postorino, Paolo and Betti, Maria Grazia and Scardamaglia, Mattia and Bittencourt, Carla and Mariani, Carlo},
	month = nov,
	year = {2018},
	pages = {768--775},
}

@article{betti_dielectric_2023,
	title = {Dielectric response and excitations of hydrogenated free-standing graphene},
	volume = {12},
	issn = {26670569},
	url = {https://linkinghub.elsevier.com/retrieve/pii/S2667056923000299},
	doi = {10.1016/j.cartre.2023.100274},
	language = {en},
	urldate = {2026-02-06},
	journal = {Carbon Trends},
	author = {Betti, Maria Grazia and Marchiani, Dario and Tonelli, Andrea and Sbroscia, Marco and Blundo, Elena and De Luca, Marta and Polimeni, Antonio and Frisenda, Riccardo and Mariani, Carlo and Jeong, Samuel and Ito, Yoshikazu and Cavani, Nicola and Biagi, Roberto and Gillespie, Peter N.O. and Hernandez Bertran, Michael A. and Bonacci, Miki and Molinari, Elisa and De Renzi, Valentina and Prezzi, Deborah},
	month = sep,
	year = {2023},
	pages = {100274},
}

@article{di_bernardo_two-dimensional_2017,
	title = {Two-{Dimensional} {Hallmark} of {Highly} {Interconnected} {Three}-{Dimensional} {Nanoporous} {Graphene}},
	volume = {2},
	copyright = {http://pubs.acs.org/page/policy/authorchoice\_termsofuse.html},
	issn = {2470-1343, 2470-1343},
	url = {https://pubs.acs.org/doi/10.1021/acsomega.7b00706},
	doi = {10.1021/acsomega.7b00706},
	language = {en},
	number = {7},
	urldate = {2026-02-06},
	journal = {ACS Omega},
	author = {Di Bernardo, Iolanda and Avvisati, Giulia and Mariani, Carlo and Motta, Nunzio and Chen, Chaoyu and Avila, Jos{\'e} and Asensio, Maria Carmen and Lupi, Stefano and Ito, Yoshikazu and Chen, Mingwei and Fujita, Takeshi and Betti, Maria Grazia},
	month = jul,
	year = {2017},
	pages = {3691--3697},
}

@article{di_bernardo_topology_2018,
	title = {Topology and doping effects in three-dimensional nanoporous graphene},
	volume = {131},
	issn = {00086223},
	url = {https://linkinghub.elsevier.com/retrieve/pii/S000862231830085X},
	doi = {10.1016/j.carbon.2018.01.076},
	language = {en},
	urldate = {2026-02-06},
	journal = {Carbon},
	author = {Di Bernardo, Iolanda and Avvisati, Giulia and Chen, Chaoyu and Avila, Jos{\'e} and Asensio, Maria Carmen and Hu, Kailong and Ito, Yoshikazu and Hines, Peter and Lipton-Duffin, Josh and Rintoul, Llew and Motta, Nunzio and Mariani, Carlo and Betti, Maria Grazia},
	month = may,
	year = {2018},
	pages = {258--265},
}

@article{kleiner_curvature_2001,
	title = {Curvature, hybridization, and {STM} images of carbon nanotubes},
	volume = {64},
	copyright = {http://link.aps.org/licenses/aps-default-license},
	issn = {0163-1829, 1095-3795},
	url = {https://link.aps.org/doi/10.1103/PhysRevB.64.113402},
	doi = {10.1103/PhysRevB.64.113402},
	language = {en},
	number = {11},
	urldate = {2026-02-06},
	journal = {Phys. Rev. B},
	author = {Kleiner, Alex and Eggert, Sebastian},
	month = aug,
	year = {2001},
	pages = {113402},
}

@article{barinov_imaging_2009,
	title = {Imaging and {Spectroscopy} of {Multiwalled} {Carbon} {Nanotubes} during {Oxidation}: {Defects} and {Oxygen} {Bonding}},
	volume = {21},
	copyright = {http://onlinelibrary.wiley.com/termsAndConditions\#vor},
	issn = {0935-9648, 1521-4095},
	shorttitle = {Imaging and {Spectroscopy} of {Multiwalled} {Carbon} {Nanotubes} during {Oxidation}},
	url = {https://advanced.onlinelibrary.wiley.com/doi/10.1002/adma.200803003},
	doi = {10.1002/adma.200803003},
	language = {en},
	number = {19},
	urldate = {2026-02-06},
	journal = {Advanced Materials},
	author = {Barinov, Alexei and Gregoratti, Luca and Dudin, Pavel and La Rosa, Salvatore and Kiskinova, Maya},
	month = may,
	year = {2009},
	pages = {1916--1920},
}

@article{apponi_transmission_2024,
	title = {Transmission through graphene of electrons in the 30 {\textendash} 900 {eV} range},
	volume = {216},
	issn = {00086223},
	url = {https://linkinghub.elsevier.com/retrieve/pii/S0008622323007479},
	doi = {10.1016/j.carbon.2023.118502},
	language = {en},
	urldate = {2026-02-06},
	journal = {Carbon},
	author = {Apponi, Alice and Convertino, Domenica and Mishra, Neeraj and Coletti, Camilla and Iodice, Mauro and Frasconi, Franco and Pilo, Federico and Blaj, Narcis Silviu and Paoloni, Daniele and Rago, Ilaria and De Bellis, Giovanni and Cavoto, Gianluca and Ruocco, Alessandro},
	month = jan,
	year = {2024},
	pages = {118502},
}

@article{balog_controlling_2013,
	title = {Controlling {Hydrogenation} of {Graphene} on {Ir}(111)},
	volume = {7},
	issn = {1936-0851, 1936-086X},
	url = {https://pubs.acs.org/doi/10.1021/nn400780x},
	doi = {10.1021/nn400780x},
	language = {en},
	number = {5},
	urldate = {2026-02-06},
	journal = {ACS Nano},
	author = {Balog, Richard and Andersen, Mie and J{\o}rgensen, Bjarke and Sljivancanin, Zeljko and Hammer, Bj{\o}rk and Baraldi, Alessandro and Larciprete, Rosanna and Hofmann, Philip and Hornek{\ae}r, Liv and Lizzit, Silvano},
	month = may,
	year = {2013},
	pages = {3823--3832},
}

@article{abdelnabi_towards_2021,
	title = {Towards free-standing graphane: atomic hydrogen and deuterium bonding to nano-porous graphene},
	volume = {32},
	issn = {0957-4484, 1361-6528},
	shorttitle = {Towards free-standing graphane},
	url = {https://iopscience.iop.org/article/10.1088/1361-6528/abbe56},
	doi = {10.1088/1361-6528/abbe56},
	number = {3},
	urldate = {2026-02-06},
	journal = {Nanotechnology},
	author = {Abdelnabi, Mahmoud Mohamed Saad and Blundo, Elena and Betti, Maria Grazia and Cavoto, Gianluca and Placidi, Ernesto and Polimeni, Antonio and Ruocco, Alessandro and Hu, Kailong and Ito, Yoshikazu and Mariani, Carlo},
	month = jan,
	year = {2021},
	pages = {035707},
}

@article{PAOLONI2023122322,
title = {Cu-phthalocyanine long-range ordered bulk growth due to the weak interaction with highly oriented pyrolytic graphite substrate},
journal = {Surface Science},
volume = {735},
pages = {122322},
year = {2023},
issn = {0039-6028},
doi = {https://doi.org/10.1016/j.susc.2023.122322},
url = {https://www.sciencedirect.com/science/article/pii/S0039602823000754},
author = {Daniele Paoloni and Alessandro Ruocco}
}

@misc{ritarossi,
      title={Graphene lattice recoil in hard X-ray photoemission: Experiment and Theory}, 
      author={Simone Ritarossi and Alice Apponi and Orlando Castellano and José Lorenzana and Domenica Convertino and Camilla Coletti and Tien-Lin Lee and Francesco Offi and Alessandro Ruocco},
      year={2026},
      eprint={2605.12330},
      archivePrefix={arXiv},
      primaryClass={cond-mat.mtrl-sci},
      url={https://arxiv.org/abs/2605.12330}, 
}

@article{nikitin_hydrogen_2008,
	title = {Hydrogen {Storage} in {Carbon} {Nanotubes} through the {Formation} of {Stable} {C}-{H} {Bonds}},
	volume = {8},
	issn = {1530-6984, 1530-6992},
	url = {https://pubs.acs.org/doi/10.1021/nl072325k},
	doi = {10.1021/nl072325k},
	language = {en},
	number = {1},
	urldate = {2026-02-09},
	journal = {Nano Lett.},
	author = {Nikitin, Anton and Li, Xiaolin and Zhang, Zhiyong and Ogasawara, Hirohito and Dai, Hongjie and Nilsson, Anders},
	month = jan,
	year = {2008},
	pages = {162--167},
}

@article{boukhvalov_hydrogen_2008,
	title = {Hydrogen on graphene: {Electronic} structure, total energy, structural distortions and magnetism from first-principles calculations},
	volume = {77},
	copyright = {http://link.aps.org/licenses/aps-default-license},
	issn = {1098-0121, 1550-235X},
	shorttitle = {Hydrogen on graphene},
	url = {https://link.aps.org/doi/10.1103/PhysRevB.77.035427},
	doi = {10.1103/PhysRevB.77.035427},
	language = {en},
	number = {3},
	urldate = {2026-02-10},
	journal = {Phys. Rev. B},
	author = {Boukhvalov, D. W. and Katsnelson, M. I. and Lichtenstein, A. I.},
	month = jan,
	year = {2008},
	pages = {035427},
}

@article{ruffieux_hydrogen_2002,
	title = {Hydrogen adsorption on sp 2 -bonded carbon: {Influence} of the local curvature},
	volume = {66},
	copyright = {http://link.aps.org/licenses/aps-default-license},
	issn = {0163-1829, 1095-3795},
	shorttitle = {Hydrogen adsorption on sp 2 -bonded carbon},
	url = {https://link.aps.org/doi/10.1103/PhysRevB.66.245416},
	doi = {10.1103/PhysRevB.66.245416},
	language = {en},
	number = {24},
	urldate = {2026-02-10},
	journal = {Phys. Rev. B},
	author = {Ruffieux, P. and Gr{\"o}ning, O. and Bielmann, M. and Mauron, P. and Schlapbach, L. and Gr{\"o}ning, P.},
	month = dec,
	year = {2002},
	pages = {245416},
}

@article{sacchi_quantifying_2005,
	title = {Quantifying the effective attenuation length in high-energy photoemission experiments},
	volume = {71},
	copyright = {http://link.aps.org/licenses/aps-default-license},
	issn = {1098-0121, 1550-235X},
	url = {https://link.aps.org/doi/10.1103/PhysRevB.71.155117},
	doi = {10.1103/PhysRevB.71.155117},
	language = {en},
	number = {15},
	urldate = {2026-02-19},
	journal = {Phys. Rev. B},
	author = {Sacchi, Maurizio and Offi, Francesco and Torelli, Piero and Fondacaro, Andrea and Spezzani, Carlo and Cautero, Marco and Cautero, Giuseppe and Huotari, Simo and Grioni, Marco and Delaunay, Renaud and Fabrizioli, Mauro and Vank{\'o}, Gy{\"o}rgy and Monaco, Giulio and Paolicelli, Guido and Stefani, Giovanni and Panaccione, Giancarlo},
	month = apr,
	year = {2005},
	pages = {155117},
}

@article{baskin_lattice_1955,
	title = {Lattice {Constants} of {Graphite} at {Low} {Temperatures}},
	volume = {100},
	copyright = {http://link.aps.org/licenses/aps-default-license},
	issn = {0031-899X},
	url = {https://link.aps.org/doi/10.1103/PhysRev.100.544},
	doi = {10.1103/PhysRev.100.544},
	language = {en},
	number = {2},
	urldate = {2026-03-01},
	journal = {Phys. Rev.},
	author = {Baskin, Y. and Meyer, L.},
	month = oct,
	year = {1955},
	pages = {544--544},
}

@article{gorodetskiy_hydrogen_2020,
	title = {Hydrogen {Plasma} {Treatment} of {Aligned} {Multi}-{Walled} {Carbon} {Nanotube} {Arrays} for {Improvement} of {Field} {Emission} {Properties}},
	volume = {13},
	issn = {1996-1944},
	url = {https://www.mdpi.com/1996-1944/13/19/4420},
	doi = {10.3390/ma13194420},
	language = {en},
	number = {19},
	urldate = {2026-03-01},
	journal = {Materials},
	author = {Gorodetskiy, Dmitriy V. and Gusel{\textquoteright}nikov, Artem V. and Kurenya, Alexander G. and Smirnov, Dmitry A. and Bulusheva, Lyubov G. and Okotrub, Alexander V.},
	month = oct,
	year = {2020},
	pages = {4420},
    }

\end{document}